\documentclass[12pt]{iopart}

\usepackage{graphicx}
\usepackage{amssymb}
\usepackage{color}

\begin{document}

\title[Local energy and power for many-particle quantum systems]{Local energy and power for many-particle quantum systems driven by an external electrical field}

\author{Guillermo Albareda}
\address{Institut de Qu\'imica Te\`orica i Computacional and Departament de Qu\'imica F\'isica, Universitat de Barcelona, Barcelona, Spain}
\ead{albareda@ub.edu}
\author{Fabio Lorenzo Traversa}
\address{Department of Physics, University of California, San Diego, La Jolla, CA 92093-0319, USA}
\ead{ftraversa@physics.ucsd.edu}
% \author{Peter H\"anggi}
% \address{Institut f\"ur Physik, Universität Augsburg, Universit\"atsstr. 1, D-86135 Augsburg, Germany}
% \ead{hanggi@physik.uni-augsburg.de}
\author{Xavier Oriols}
\address{Depertament d'Enginyeria Electr\`onica, Universitat Aut\`onoma de Barcelona, Bellaterra, Spain}
\ead{xavier.oriols@uab.cat}
\vspace{10pt}
\begin{indented}
\item[]January  2016
\end{indented}

\begin{abstract}
We derive expressions for the expectation values of the local energy and the local power transferred by an external electrical field to a many-particle system of interacting spinless electrons. 
In analogy with the definition of the (local) presence and current probability densities, we construct a local energy operator such that the time-rate of change of its expectation value provides information on the spatial distribution of power. 
Results are presented as functions of an arbitrarily small volume $\Omega$, and physical insights are discussed by means of the quantum hydrodynamical representation of the wavefunction, which is proven 
to allow for a clear-cut separation into contributions with and without classical correspondence. 
Quantum features of the local power are mainly manifested through the presence of non-local sources/sinks of power and through the action of forces with no classical counterpart. 
Many-particle classical-like effects arise in the form of current-force correlations and through the inflow/outflow of energy across the boundaries of the volume $\Omega$.
Interestingly, such intriguing features are only reflected in the expression for the local power when the volume $\Omega$ is finite. 
Otherwise, for closed systems with $\Omega \to \infty$,  we recover a classical-like single-particle expression.
\end{abstract}

% Uncomment for PACS numbers

% Uncomment for keywords
\vspace{2pc}

% Uncomment for Submitted to journal title message
%\submitto{\JSTAT}

% Uncomment if a separate title page is required
%\maketitle
 
% For two-column output uncomment the next line and choose [10pt] rather than [12pt] in the \documentclass declaration
%\ioptwocol

\section{Introduction}
Conservation laws are extraordinarily useful mathematical tools for determining the time-evolution of complex systems. 
Among them, energy conservation is perhaps the most conspicuous one. 
Different types of energies (kinetic, potential, chemical, etc.) are commonly defined in order to identify an expression for a global property of the system that do not change in time, and hence 
use it to impose restrictions on the equations of motion of complex systems.

For open (non-isolated) systems interacting with an environment out-of-equilibrium, energy, although not preserved, remains a valid and useful quantity. 
The notion of power is introduced in this context as a measure of the rhythm at which systems gain or loose energy through the interaction with the environment.  
From a theoretical point of view, the concept of power has been used to explore the extension of thermodynamic laws to the realm of quantum mechanics~\cite{thermo1,thermo2,thermo3,thermo4,thermo5}. 
From an operational viewpoint, power plays a central role, e.g., in evaluating the performance of emergent electronic and nanomechanical devices~\cite{electronics1,electronics2,electronics3,electronics4,electronics5}, for which low-power consumption is a prevailing requirement. 

For quantum systems that preserve the number of particles, the power supplied by an external (driving or dissipative) force is defined as the time-rate of change of the expectation value of 
the internal energy~\cite{yang1}. 
For quantum systems which are open to the flow of particles, however, energy can turn into a fuzzy concept unless it is carefully redefined.  
As will be shown here, the evaluation of the spatial distribution of energy and power (defined for small regions of the physical space) requires a careful theoretical approach. 

Access to spatially resolved energy and power can be relevant in the context of local control of particle heating or cooling through the action of laser fields~\cite{cooling}.
But more importantly, this information can be used to assess the performance of electronic devices, for which determining the spatial distribution of energy and power along the source, gate or drain regions, is of paramount 
importance~\cite{local_electrons1,local_electrons2,local_electrons3}. 
In classical mechanics, the access to local energy and power information does not pose any mathematical or conceptual difficulty~\cite{yang_classic}.  
However, in quantum mechanics, although not far from being measurable~\cite{measuring}, the evaluation of such spatially resolved information is accompanied by some conceptual and mathematical intricacies. 

In this work we address the specific question of what are the spatial distributions of energy and power transferred by an external electrical field to an ensemble of electrons.  
In particular, we want to provide a \textit{mechanistic} answer to this question without resorting to any kind of thermodynamical argument. 
To this end, we first define a local energy operator whose expectation value provides a measure of the energy enclosed in an arbitrarily small volume $\Omega$ of the physical space. 
Local power will be then defined as the time-rate of change of this expectation value.
Significantly, we will show that the exact expression for the local power depends on intriguing many-particle current-force correlations and non-local quantum terms that are not present in the standard equation 
for the total (spatially integrated) electrical power~\cite{yang2}.

After this introduction, in section II we describe the general system under study and motivate the use of local operators. 
Then, in section III, we define the local energy operator and evaluate its expectation value. In section IV, we identify the local power as the time derivative of the local energy and discuss its physical soundness. 
We conclude in section V. 
  
\section{Preliminary discussion}
\label{sec2}
Before deriving expressions for the local energy and power, we provide here a comprehensive description of the system (and environment) that will be approached in this work. 
We also introduce here the concept of local operator, which will be used later in our derivations.

We consider first a quantum system which is kept out of equilibrium by the action of an external (effective) electric field. This system can exchange energy but not particles with its environment.
Later we will define a smaller volume $\Omega$, enclosed in the first one, which can exchange both energy and particles with its surroundings. 
This volume can be made as small as required, and thus we will call any observable associated to it \emph{local}. 
Our main goal in this work is to find an expression for the power and energy associated to this second volume $\Omega$. 

\subsection{Energy and power for many-electrons driven by an external electrical field} 
\label{sec2.1} 
Consider an ensemble of interacting spinless electrons under the action of an external electrical field and defined by the $N$-particles state $|\psi(t)\rangle$. 
This state can be written in the position representation as the many-particle (scalar) wavefunction $\psi(\mathbf{r},t)=\langle \mathbf{r}|\psi(t)\rangle$, 
where the ket $|\mathbf{r} \rangle= |\mathbf{r}_1 \rangle \otimes...\otimes |\mathbf{r}_{N} \rangle$ collectively denotes the position of the electrons in the $3N$-dimensional configuration space (and $\otimes$ represents the direct product). 
We consider here all particles to be spinless electrons. 
The generalization of our results for multi-component (vector) spinors does not add any conceptual complexity but it certainly complicates our mathematical derivation. 
Throughout this work we use atomic units ($m_{e}=1$, $e=1$, $\hbar =1$ and Coulomb's constant $k_{e} =1$) and bold letters indicate vectors defined in the 3-dimensional Cartesian space, 
i.e. $|\mathbf{r}_k \rangle=|x_k \rangle \otimes |y_k\rangle \otimes |z_k \rangle$.
We assume that state $|\psi(t)\rangle$ is effectively governed by the following time-dependent Schr\"odinger equation:
\begin{equation}
i\frac{\partial}{\partial t}|\psi(t)\rangle = \hat H(t)   |\psi(t)\rangle = \left(  \hat K + \hat U(t) \right)  |\psi(t)\rangle,
\label{schrodinger}%
\end{equation}
where $\hat K = \frac{1}{2}\sum_k^N \hat \mathbf{p}_k \cdot \hat \mathbf{p}_k$ is the many-body kinetic energy operator, with $\hat \mathbf{p}_k$ the linear momentum operator. 
The term $\hat U(t) = \hat U_{cou} + \hat U_{ext}(t)$ represents a (scalar) potential energy operator that accounts for the Coulomb interaction $\hat U_{cou}$ among the $N$ electrons and also for
their interaction with an external electrical field through $\hat U_{ext}(t)$. 
In the position representation it reads:
\begin{equation}
{U}(\mathbf{r},t) = {U}_{cou}(\mathbf{r}) + \sum_{k}^{N} {U}_{ext}(\mathbf{r}_k,t).
\label{poten}
\end{equation} 
The Coulomb interaction among electrons can be written in terms of the two-particle operator ${U}_{k,j} \equiv {U}_{k,j}(\mathbf{r}_k-\mathbf{r}_j)$ as:
\begin{equation}
{U}_{cou}(\mathbf{r}) = \sum_{k}^{N} \sum_{j>k}^{N} {U}_{k,j}(\mathbf{r}_k-\mathbf{r}_j).
\label{poten2}
\end{equation} 
Contrarily, the external potential ${U}_{ext}(\mathbf{r}_k,t)$ represents a single-particle (effective) operator that accounts for, e.g., the bias generated by an external battery, the effect of an external laser field (in the dipole approximation), 
or the time-dependent potential generated by quasi-static nuclei (in the Born-Oppenheimer approximation).
In the separation of the potential energy operator into Coulombic and external components, it is implicitly assumed that there are particles other than the $N$ electrons whose effect on the evolution of the state 
$|\psi(t)\rangle$ can be effectively modeled through a single-particle field ${U}_{ext}(\mathbf{r}_k,t)$.
This separation is commonly used in the literature and, as discussed bellow, it plays a crucial role in the proper definition of the concepts of energy and power for open systems.

Let us now address the question of what is the energy associated with the system described by the state $|\psi(t)\rangle$ (and obeying the effective Hamiltonian $\hat H(t)$ in Eq.~\eref{schrodinger}).
To correctly answer this question we must guarantee the following three requirements~\cite{yang_classic}:
\begin{itemize}
 \item For an isolated (conservative) system the energy must be conserved, i.e., $\langle \hat{E}_{isolated} \rangle = constant$.
 \item If the system is not isolated, however, but is in an external time-dependent field which supplies power $P(t)$, the definition of the energy and the power of the system must satisfy ${ d\langle \hat{E}(t) \rangle}/{dt} = P(t)$. 
 \item When the power supplied by the external field is zero, $P(t)=0$, then $\langle \hat{E}(t) \rangle = \langle \hat{E}_{isolated} \rangle$. 
\end{itemize}
Such very simple arguments allows us to unequivocally define the energy (or \emph{internal} energy if preferred) associated to the $N$ electrons described by Eq.~\eref{schrodinger} as:
\begin{equation}
\fl
\langle \hat{E}(t) \rangle = \langle \psi |\left( \hat H - \hat U_{ext}(t) \right) | \psi \rangle =\sum_{k}^N \int_\infty d\mathbf{r} \;  \psi^*(\mathbf{r},t) \left( \frac{1}{2} \sum_k^N \nabla_k^2 + U_{cou}(\mathbf{r}) \right)  \psi(\mathbf{r},t).
\label{energy3}%
\end{equation}
The term $\frac{1}{2} \sum_k^N \nabla_k^2$ is the many-particle kinetic energy operator in the position representation, with $\nabla^2_{k}=\nabla_{k} \cdot \nabla_{k}$ being the laplacian operator 
and $\nabla_{k}=\mathbf{u}_{x}\;\frac{\partial}{\partial x_k}+\mathbf{u}_{y}\;\frac{\partial}{\partial y_k}+\mathbf{u}_{z}\;\frac{\partial}{\partial z_k}$ the nabla operator, 
where $\{\mathbf{u}_{x},\mathbf{u}_{y},\mathbf{u}_{z}\}$ are unitary vectors pointing respectively in the three directions of the physical space. 
Hereafter, we also use the definition $d\mathbf{r}=dx_1 \otimes dy_1 \otimes dz_1 \otimes.....\otimes dx_N \otimes dy_N \otimes dz_N$.

To see that Eq. \eref{energy3} represents the energy of the system, we only need to realize that the electrical power $P(t)$ provided by the external field, ${U}_{ext}(t)$, 
can be identified with the time-rate of change of the expectation value of the energy defined in Eq.~\eref{energy3}, i.e.:
\begin{eqnarray}
 P(t) = \frac{d}{dt}\langle \hat{E}(t) \rangle &=& i \langle  \left[ \hat H, \hat K \right]  \rangle  +  i \langle  \left[ \hat H, \hat U_{cou} \right]  \rangle \nonumber\\
      &=& i \langle  \left[ \hat U(t), \hat K \right]  \rangle + i \langle  \left[ \hat K, \hat U_{cou} \right] + i \langle  \left[ \hat U(t), \hat U_{cou} \right]  \rangle   \nonumber\\
      &=& i \langle  \left[ \hat U_{ext}(t), \hat K \right]  \rangle,
 \label{total_power0}
\end{eqnarray}
where in the last equality we have used that the operators $\hat U(t)$ and $\hat U_{cou}$ commute. 
The commutator in the last equality of Eq.~\eref{total_power0} (in the position representation) gives:
\begin{eqnarray}
\fl
 P(t) &=&  \frac{i}{2} \sum_k^N \int_\infty d\mathbf{r} \left( |\Psi|^2  \nabla_k^2  U_{ext}    +   2\Psi^* (\nabla_k  U_{ext}) (\nabla_k \Psi) \right)   \nonumber\\
\fl   
   &=& \frac{i}{2} \sum_k^N \int_\infty d\mathbf{r} \left(  \nabla_k \left(  |\Psi|^2\nabla_k  U_{ext}  \right)   -  (\Psi\nabla_k \Psi^*)(\nabla_k U_{ext})  +  (\Psi^*\nabla_k \Psi)(\nabla_k U_{ext})  \right)  \nonumber\\
\fl   
   &=& \sum_k^N \int_\infty d\mathbf{r}    \mathbf{J}_k \cdot \mathbf{F}^{ext},
\label{total_power}         
\end{eqnarray}
where in the last equality we used the Gauss theorem and we have defined the external force $\mathbf{F}^{ext}(\mathbf{r}_k,t) = - \nabla_k U_{ext}(\mathbf{r}_k,t)$ and the $k$-th component of the 
standard probability current density $\mathbf{J}_k = \langle \hat \mathbf{J}_k \rangle = \frac{-i}{2}(\Psi^*\nabla_k\Psi - \Psi\nabla_k\Psi^*)$~\cite{cohen}. 
Notice that Eq.~\eref{total_power} corresponds to the standard definition of electrical power (see Eq.~(3.6) in Ref.~\cite{yang2}).

The above discussion brings us to the main question of this work: what is the spatial distribution of the electrical energy and power $P(t)$ found in Eqs.~\eref{energy3} and \eref{total_power}? 
As we will show in the remaining part of this section, answering this questions requires the use of local energy operators.

\subsection{On the physical meaning of local operators}\label{phys_meaning}
\begin{figure}
\includegraphics{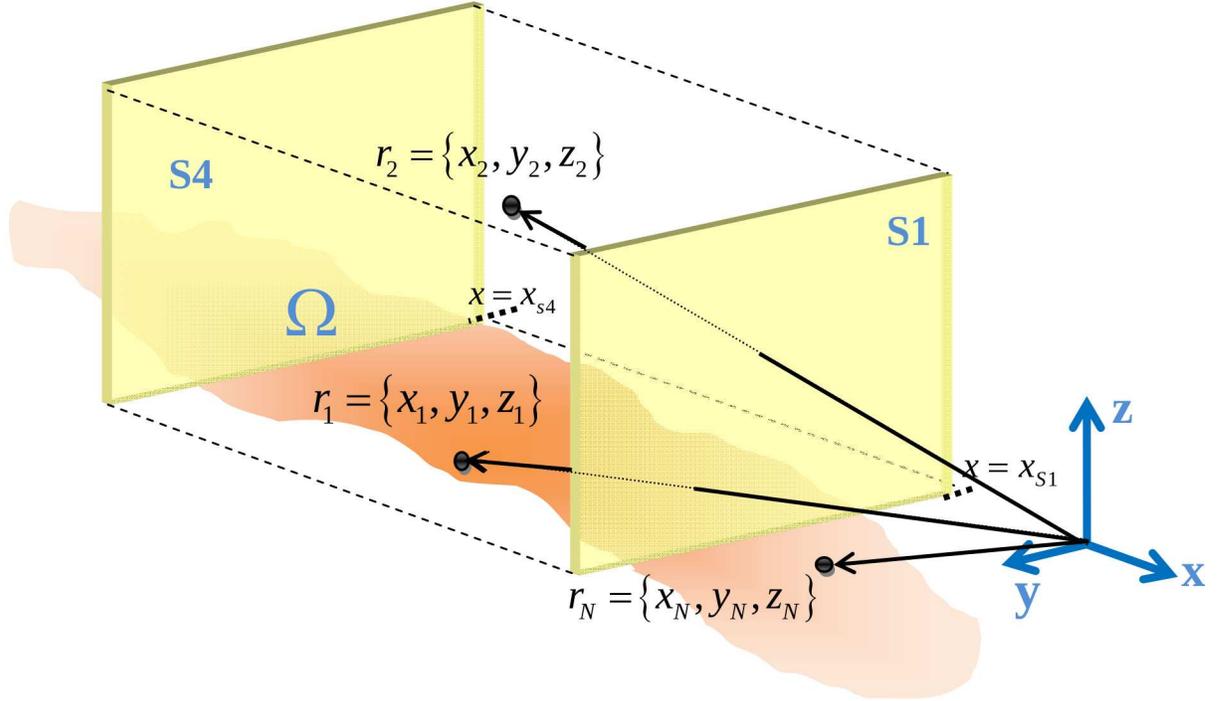}%height=6.75cm
\caption{
Schematic representation of the open volume $\Omega$ in which the local energy and power are calculated. The volume $\Omega$ interchanges energy and particles with outside. For simplicity, we consider the surfaces $S1$ and $S4$ open to the transit of particles, while in the rest of surfaces a vanishing wavefunction is cosnidered.
The total number of particles inside and outside $\Omega$ is $N$, but only some of them  contribute to the energy and power inside $\Omega$. }
\label{scheme}
\end{figure}

Consider an (arbitrarily) small volume $\Omega$ (depicted in Fig. \ref{scheme}), which, for simplicity, has only two surfaces, $S4$ and $S1$, open to the transit of particles. 
Please note that making all surfaces open to the flux of particles would make the notation very tedious without proving any additional insight. 
Given this volume $\Omega$, we address now three very simple problems that will be proved to be helpful later in the derivation of our main results.

Let us first compute the probability of finding a particle $k$ inside the volume $\Omega$ (irrespective of where the rest of $N-1$ particles are). 
According to Born rules, the probability of finding the $N$ particles at positions $\mathbf{r} = (\mathbf{r}_1,...,\mathbf{r}_{N})$ in the configuration space is 
$\langle \psi  | \hat \Pi(\mathbf{r}) | \psi \rangle  =|\psi(\mathbf{r},t)|^2$, where $\hat \Pi(\mathbf{r})=|{\mathbf{r}}\rangle  \langle {\mathbf{r}}|$ is the position operator.  
Therefore, we can try to answer the question above by introducing a similar (position-like) operator:
 \begin{eqnarray}
\hat \Pi_{k,\Omega}=\int_\infty d\bar{\mathbf{r}}_k \int_\Omega d\mathbf{r}_k |\bar{\mathbf{r}}_k\rangle  \langle \bar{\mathbf{r}}_k|\otimes |\mathbf{r}_k\rangle \langle \mathbf{r}_k| = \bar{\mathbf{1}}_k \otimes  \int_\Omega d\mathbf{r}_k   |\mathbf{r}_k\rangle \langle \mathbf{r}_k|,
\label{where1}
\end{eqnarray} 
where we have defined  $\bar{\mathbf{r}}_{k} = (\mathbf{r}_1,..,\mathbf{r}_{k-1},\mathbf{r}_{k+1},..,\mathbf{r}_N)$,
$|\bar{\mathbf{r}}_k\rangle  \langle \bar{\mathbf{r}}_k|=|\mathbf{r}_1\rangle  \langle \mathbf{r}_1|\otimes..\otimes|\mathbf{r}_{k-1}\rangle  \langle \mathbf{r}_{k-1}|\otimes|\mathbf{r}_{k+1}\rangle  \langle \mathbf{r}_{k+1}|\otimes..\otimes |\mathbf{r}_N\rangle  \langle \mathbf{r}_N|$,
and $\bar{\mathbf{1}}_k=\int_\infty d\bar{\mathbf{r}}_k |\bar{\mathbf{r}}_k\rangle  \langle \bar{\mathbf{r}}_k|$.
The expectation value of the operator in \eref{where1} provides the required information: 
\begin{eqnarray}
\langle  \hat \Pi_{k,\Omega}(t) \rangle=  \int_\infty d\bar{\mathbf{r}}_k \int_\Omega d\mathbf{r}_k |\psi(\mathbf{r},t)|^2.
\label{where2}
\end{eqnarray}
Notice that the above expectation value is valid either for fermions or bosons. 
For identical particles, the shape of the wave function imposes that $ \langle \hat \Pi_{k,\Omega}(t) \rangle =\langle \hat  \Pi_{j,\Omega}(t) \rangle$ for any $j \neq k$, and thus particles cannot be distinguish one from each other.

We now move to a second illustrating example. We want to compute the expectation value of the $k$-th component of the probability current density given that the $k$-th electron is sitting inside the volume $\Omega$ (independently of the location of the rest of electrons). 
We know that the current density operator can be written as $\hat \mathbf{J}_k(\mathbf{r})=\frac{1}{2} \left(|\bar{\mathbf{r}}_k\rangle  \langle \bar{\mathbf{r}}_k|\otimes |\mathbf{r}_k\rangle \langle \mathbf{r}_k| \hat\mathbf{p}_k+ |\bar{\mathbf{r}}_k\rangle \langle \bar{\mathbf{r}}_k| \otimes \hat\mathbf{p}_k  | \mathbf{r}_k\rangle \langle \mathbf{r}_k| \right)$
and its expectation value simply reads  $\langle  \hat \mathbf{J}_k(\mathbf{r},t) \rangle = \frac{-i}{2} \Big( \psi^* {\nabla_k \psi} -\psi \nabla_k \psi^*\Big)$~\cite{cohen}.  
Similarly, we can define the following local current operator:
\begin{eqnarray}
 \hat \mathbf{J}_{k,\Omega} &=& \frac{1}{2} \int_\infty d\bar{\mathbf{r}}_k \int_\Omega d\mathbf{r}_k  |\bar{\mathbf{r}}_k\rangle  \langle \bar{\mathbf{r}}_k|\otimes \Big( |\mathbf{r}_k\rangle \langle \mathbf{r}_k| \hat\mathbf{p}_k+ \hat\mathbf{p}_k |\mathbf{r}_k\rangle \langle \mathbf{r}_k|,\Big)\nonumber\\
 &=& \bar{\mathbf{1}}_k \otimes \frac{1}{2} \int_\Omega d\mathbf{r}_k  \Big( |\mathbf{r}_k\rangle \langle \mathbf{r}_k| \hat\mathbf{p}_k+ \hat\mathbf{p}_k |\mathbf{r}_k\rangle \langle \mathbf{r}_k|,\Big) ,
\label{current2}%
\end{eqnarray}
whose expectation value provides the requested information, i.e.:
\begin{equation}
 \langle \hat \mathbf{J}_{k,\Omega} \rangle = \frac{-i}{2} \int_\infty d\bar{\mathbf{r}}_k \int_\Omega d\mathbf{r}_k  \Big( \psi^*(\mathbf{r},t) {\nabla_k \psi(\mathbf{r},t)} -\psi(\mathbf{r},t) \nabla_k \psi^*(\mathbf{r},t)\Big).
\end{equation}

Finally, in a third example, we discuss the operator that provides the probability of finding two particles, say the $k$-th and $j$-th electrons, inside the volume $\Omega$ no matter where the other electrons are.
We define such an operator as:
\begin{eqnarray}
\hat \Pi_{k,j,\Omega} &=& \int_\infty d\bar{\mathbf{r}}_{k,j} \int_\Omega d\mathbf{r}_k \int_\Omega d\mathbf{r}_j |\bar{\mathbf{r}}_{k,j}\rangle  \langle \bar{\mathbf{r}}_{k,j}| \otimes |\mathbf{r}_k\rangle \langle \mathbf{r}_k| \otimes |\mathbf{r}_j\rangle \langle \mathbf{r}_j| 
   \nonumber\\
   &=& \bar{\mathbf{1}}_{k,j} \otimes  \int_\Omega d\mathbf{r}_{k,j} \int_\Omega d\mathbf{r}_j  |\mathbf{r}_k\rangle \langle \mathbf{r}_k| \otimes |\mathbf{r}_j\rangle \langle \mathbf{r}_j|,
   \label{where_local}
\end{eqnarray} 
where we have now defined $\bar{\mathbf{r}}_{k,j} = (\mathbf{r}_1,..,\mathbf{r}_{k-1},\mathbf{r}_{k+1},..,\mathbf{r}_{j-1},\mathbf{r}_{j+1},..,\mathbf{r}_N)$,
$ |\bar{\mathbf{r}}_{k,j}\rangle  \langle \bar{\mathbf{r}}_k| = ..\otimes|\mathbf{r}_{k-1}\rangle  \langle \mathbf{r}_{k-1}|\otimes|\mathbf{r}_{k+1}\rangle  \langle \mathbf{r}_{k+1}|\otimes..
      \otimes|\mathbf{r}_{j-1}\rangle  \langle \mathbf{r}_{j-1}|\otimes|\mathbf{r}_{j+1}\rangle  \langle \mathbf{r}_{j+1}|\otimes..
$
and $\bar{\mathbf{1}}_{k,j} = \int_\infty d\bar{\mathbf{r}}_{k,j} |\bar{\mathbf{r}}_{k,j}\rangle  \langle \bar{\mathbf{r}}_{k,j}|$.
The expectation value of the operator in Eq.~\eref{where_local} provides the wanted information, i.e.:
\begin{eqnarray}
\langle  \hat \Pi_{k,j,\Omega}(t) \rangle=  \int_\infty d\bar{\mathbf{r}}_{k,j} \int_\Omega d\mathbf{r}_k \int_\Omega d\mathbf{r}_j \; |\psi(\mathbf{r},t)|^2.
\label{where2}
\end{eqnarray}

The above three examples will be very helpful in our definition of the local energy. In particular, the last example above will be relevant in the definition of the internal (two-particle) Coulombic operator.

\section{Spatial distribution of energy \label{kinetic_energy}}
We are now in a position to derive an expression for the local energy operator. 
Let us first define such an operator, generically, as the sum of a local kinetic energy operator and a local potential energy operator, i.e.: $\hat E_{\Omega} = \hat K_{\Omega} +  \hat C_{\Omega}$.
Bellow we discuss separately these two operators $\hat K_{\Omega}$ and $\hat C_{\Omega}$.

\subsection{Local kinetic energy operator}
Taking advantage of the example for the local current density operator in Eq.~\eref{current2}, we here define the (hermitian) local kinetic energy operator $\hat K_{\Omega}$ associated to the volume $\Omega$ as: 
\begin{eqnarray}
\hat K_\Omega 
      &=&  \frac{1}{4} \sum_k^N  \int_\infty d\bar{\mathbf{r}}_k  \int_\Omega d\mathbf{r}_k \Big( |\bar{\mathbf{r}}_k\rangle  \langle \bar{\mathbf{r}}_k|\otimes |\mathbf{r}_k\rangle \langle \mathbf{r}_k| \hat\mathbf{p}_k^2  +  |\bar{\mathbf{r}}_k\rangle \langle \bar{\mathbf{r}}_k| \otimes \hat\mathbf{p}_k^2 |\mathbf{r}_k\rangle \langle \mathbf{r}_k| \Big)\nonumber\\
            &=&  \frac{1}{4} \sum_k^N   \int_\Omega d\mathbf{r}_k \; \bar{\mathbf{1}}\otimes \Big( |\mathbf{r}_k\rangle \langle \mathbf{r}_k| \hat\mathbf{p}_k^2  + \hat\mathbf{p}_k^2 |\mathbf{r}_k\rangle \langle \mathbf{r}_k| \Big).
\label{kinetic_energy_open}
\end{eqnarray}
The above operator provides the expectation value of the kinetic energy comprised in the (two-terminal) open volume $\Omega$:
\begin{eqnarray}
  \left\langle  \hat K_\Omega \right\rangle  = Real\left(\sum_{k}^N \int_\infty d\bar{\mathbf{r}}_k \int_\Omega d\mathbf{r}_k \;  \psi^*(\mathbf{r},t) \frac{ \nabla_k^2}{2}   \psi(\mathbf{r},t)\right).
\label{previous}%
\end{eqnarray}
Equation \eref{previous} accounts for any $k$-th electron contribution to the kinetic energy in the volume $\Omega$ independently of where the rest of $N-1$ electrons are~\footnote{Because of the restricted limits of the integral in Eq.~\eref{previous}, only the real part of its outcome is considered.}.
Notice that an alternative definition of the local kinetic energy operator could have been:
$\hat K_\Omega =  \frac{1}{4} \sum_k^N  \int_\Omega d\mathbf{r}   \Big( |\mathbf{r}\rangle  \langle \mathbf{r}| \hat\mathbf{p}_k^2  +   \hat\mathbf{p}_k^2 |\mathbf{r}\rangle \langle \mathbf{r}| \Big)$.
However, such a local operator does not provide the desired information, i.e., its expectation value would be only different from zero in those situations where the support of all 1-dimensional reduced probability densities 
$\rho_k(\mathbf{r}_k,t) = \int_\infty d\bar{\mathbf{r}}_k |\Psi|^2$ simultaneously intersect the volume $\Omega$.

In view of Eq.~\eref{previous}, it is worth noting that in classical mechanics the increment of the $k$-th electron kinetic energy is always reflected into the increase of its associated electrical current. 
That is not the case in quantum mechanics, where the expectation values of the kinetic energy and current density are not directly related. 
To better appreciate this important point, following the hydrodynamic formulation of the wave function \cite{holland,bohm1,Wyatt,bohm2}, it is useful to introduce the polar expression of the wavefunction  
$\Psi(\mathbf{r},t) = R(\mathbf{r},t)e^{i S(\mathbf{r},t)}$ into Eq.~\eref{previous}  to write: 
\begin{eqnarray}
  \left\langle  \hat K_\Omega \right\rangle  = \sum_k^N \int_\infty  d\bar{\mathbf{r}}_k \int_\Omega d\mathbf{r}_k   R^2 \Big( Q_k + \frac{ (\nabla_k S)^2}{2} \Big),
\label{kineticintpo}%
\end{eqnarray}
where ${Q_k}(\mathbf{r},t) = -\nabla_k^2 R/(2R)$ is the $k$-th component of the so-called quantum potential~\cite{holland,bohm1,Wyatt,bohm2}. 
A real time-independent quantum state, i.e. with current density equal to zero,  does still contribute to Eq. \eref{kineticintpo} in the form of  $Q_k$. 
This issue will be discussed with more detail in Sec.~\ref{meaning_energy}.

\subsection{Local Coulombic energy operator}\label{local_pot}
In order to define the operator related with the potential energy associated to the volume $\Omega$, we now rely on the third example introduced in Section~\ref{phys_meaning}. 
The operator in Eq.~\eref{where_local} was designed to provide the probability of finding simultaneously two electrons $k$ and $j$ at the same time in the volume $\Omega$, irrespective of the position of the other $N-2$ electrons.
Because the Coulombic interaction is defined through a two-particle potential energy operator, $\hat U_{j,k}$, we can define the potential energy in the volume $\Omega$ in terms of the following operator:
\begin{eqnarray}
\hat C_\Omega  = \sum_k^N \sum_{j>k}^N  \int_\infty d\bar{\mathbf{r}}_{j,k}  \int_\Omega d\mathbf{r}_k \int_\Omega d\mathbf{r}_j |\mathbf{r}\rangle  \langle \mathbf{r}| \hat U_{j,k}.  
\label{Coulombint}%
\end{eqnarray}
The expectation value of the above operator represents a measure of the internal (Coulombic) energy in the volume $\Omega$, i.e.: 
\begin{eqnarray}
\langle \hat C_{\Omega}\rangle = \sum_k^N \sum_{j>k}^N  \int_\infty d\bar{\mathbf{r}}_{j,k}  \int_\Omega d\mathbf{r}_k \int_\Omega d\mathbf{r}_j  |\psi(\mathbf{r},t)|^2 U_{j,k}(\mathbf{r}_j,\mathbf{r}_k),
\label{Coulombint2}%
\end{eqnarray}
which can be written in terms of the hydrodynamic form of the wavefunction as:
\begin{eqnarray}
\langle \hat C_{\Omega}\rangle = \sum_k^N \sum_{j>k}^N  \int_\infty d\bar{\mathbf{r}}_{j,k}  \int_\Omega d\mathbf{r}_k \int_\Omega d\mathbf{r}_j  R^2 U_{j,k}(\mathbf{r}_j,\mathbf{r}_k).
\label{Coulombint3}%
\end{eqnarray}
In virtue of Eq.~\eref{Coulombint3}, only the potential energy associated to pairs of electrons lying both in the volume $\Omega$ is taken into account. In other words, the potential energy associated to 
the (Coulomb) interaction of a pair of electrons, one sitting in the volume $\Omega$ and the other one outside, is considered external, and thus, not added up into Eq.~\eref{Coulombint3}.

This particular way of defining the potential energy operator in Eq.~\eref{Coulombint} will be proved correct later. 
Here, we motivate its definition by discussing a particular example where an electronic device is biased by the action of an external electrical field. Roughly speaking, the battery providing the given bias can be understood as ultimately made of a large number of electrons. 
In fact, it is the action of all the electrons of the battery on the electrons of our system what is regarded in a practical way as the bias.
The reason why we commonly call this bias \emph{external} potential is simply because we are not considering the electrons of the battery among the simulated particles, and hence 
the (internal) energy associated to our system (e.g., the conduction electrons) does not include the energy of the battery.
Therefore, coming back to our problem, when we look at the volume $\Omega$ as our system of interest (i.e., the local region where we want to compute the energy and power), 
all electrons outside this region play the role of an additional external field that is added up to the external field $ U_{ext}(t)$ \cite{extra}.

\subsection{On the physical meaning of the local expectation value of the energy}\label{meaning_energy}
We can now gather Eqs.~\eref{kineticintpo} and \eref{Coulombint3} to write an expression for the expectation value of the local energy in the volume $\Omega$:
\begin{eqnarray}
E_{\Omega}(t) = \langle \hat E_{\Omega}\rangle &=& \sum_k^N \int_\infty  d\bar{\mathbf{r}}_k \int_\Omega d\mathbf{r}_k   R^2 \Big( Q_k + \frac{ (\nabla_k S)^2}{2} \Big) \nonumber\\ 
&+&\sum_k^N  \sum_{j>k}^N  \int_\infty d\bar{\mathbf{r}}_{j,k}  \int_\Omega d\mathbf{r}_k \int_\Omega d\mathbf{r}_j  R^2 U_{j,k} .
\label{local_energy}%
\end{eqnarray}
To gain some physical insight into the meaning of Eq.~\eref{local_energy}, we can first consider its classical limit \cite{Rosen}  by just eliminating any contribution associated to the quantum potential $Q_k$, 
and reinterpreting $\nabla_k S$ as the (local) $k$-th electron velocity field, i.e.:
\begin{eqnarray}
E_{\Omega,class}(t) &=& \sum_k^N \int_\infty  d\bar{\mathbf{r}}_k \int_\Omega d\mathbf{r}_k   R_{class}^2   \frac{ (\nabla_k S_{class})^2}{2} \nonumber\\ 
&+&\sum_k^N  \sum_{j>k}^N  \int_\infty d\bar{\mathbf{r}}_{j,k}  \int_\Omega d\mathbf{r}_k \int_\Omega d\mathbf{r}_j  R_{class}^2 U_{j,k}.
\label{local_energy_classic}%
\end{eqnarray}
According to classical statistical mechanics \cite{Rosen}, an ensemble of electrons can be described by a wavefunction $\Psi_{class} = R_{class}e^{iS_{class}}$, where the phase $S_{class}(\mathbf{r},t)$ is nothing but the action defined through the classical 
Hamilton-Jacobi equation, and $R_{class}(\mathbf{r},t)$ obeys a classical continuity equation~\cite{bohm1,Rosen}.
The classical expression in Eq.~\eref{local_energy_classic} allows us to regard the term $R^2Q_k$ in Eq.~\eref{local_energy} as an explicit signature of the quantum nature of our electronic system.
In this regard, to better understand the role played by the quantum potential $Q_k$ in Eq.~\eref{local_energy}, let us neglect the Coulomb interaction and consider a separable many-particle wavefunction
$\psi(\mathbf{r},t)=\psi_{1,x}(x_1)..\psi_{N,z}(z_N)exp(iEt)$. Then, we can gain some insight into the meaning of the term $R^2Q_k$ by considering the following two examples: 

\begin{enumerate} 
\item Let us assume first that each single-particle state can be represented through a plane wave $\psi_{k,x}(x_k)=exp(i p_{k,x} x_k)$ with momentum $p_{k,x}$. 
For a plane wave $R_{k,x}=cte$, and then $Q_{k,x}=0$. Furthermore, since $S_{k,x} = p_{k,x} x_k$, then $(\nabla_k S)^2/2=(p_{k,x}^2+p_{k,y}^2+p_{k,z}^2)/2$ and $\mathbf{J}_k =R^2( p_{k,x} \mathbf{u}_x+p_{k,y} \mathbf{u}_y+p_{k,z} \mathbf{u}_z) $. 
Therefore, in the limit where $Q_k=0$ (for any $k$), the expectation value of the kinetic energy is proportional to the current density. 

\item Alternatively, consider the single-particle states $\psi_{k,x}(x_k)$ to be eigenfunctions of a (infinite) quantum well. 
Such states do not carry current density because their associated wavefunctions are real. 
In this respect, the expectation value of the kinetic energy operator is still different from zero but now equal to the quantized energy levels that the states are occupying. 
Therefore, as shown in Eq.~\eref{kineticintpo}, it is possible to increase the kinetic energy without modifying the current density. 
Notice that this very simple example has no classical counterpart. 
\end{enumerate}

Therefore, in general, the expectation value of the (local) kinetic energy includes a classical-like contribution directly related to the current density, and a non-classical component associated to the quantum potential. 
This is a well-known result in the context of a hydrodynamical formulation of quantum mechanics  \cite{holland,bohm1,Wyatt,bohm2}.  
Finally, notice that in the limit $\Omega \to \infty$, Eq. \eref{local_energy} simply reduces to the expectation value in Eq.~\eref{energy3}.      
     
\section{Spatial distribution of power \label{power}}
We finally jump into the question of what is the spatial distribution of power. As already mentioned, such a local quantity must be consistent with the definition of the local energy in Eq.~\eref{local_energy}. 
In particular we must guarantee that:
\begin{eqnarray}
  P_\Omega(t) = \langle \hat P \rangle = \frac{d}{dt}\langle \hat E_{\Omega}\rangle.
  \label{E_P_relation}
\end{eqnarray}
We can certainly look for an operator $\hat P$ that obeys the above equation, but it seems much practical here to directly evaluate the time-derivative of the expectation value of the local energy.
This is precisely what we will do in the next subsection.

\subsection{Expectation value of local power}
The time-derivative of the expectation value of the local energy in Eq.~\eref{local_energy} reads: 
\begin{eqnarray}
 P_\Omega(t) &=& \frac{d}{dt} \langle \hat E_{\Omega}\rangle=\frac{d}{dt} \left\langle \hat K_\Omega + \hat C_{\Omega}\right\rangle \nonumber\\
  &=& \sum_k^N \int_\infty d\bar{\mathbf{r}}_k \int_\Omega d\mathbf{r}_k \frac{\partial R^2}{\partial t} \Big(Q_k + \frac{(\nabla_k S)^2}{2} \Big) \nonumber\\
  &+& \sum_k^N \int_\infty d\bar{\mathbf{r}}_k \int_\Omega d\mathbf{r}_k R^2\Big(  \frac{\partial Q_k}{\partial t} +  (\nabla_k S) \cdot \Big(\nabla_k \frac{\partial S}{\partial t}\Big) \Big)  \nonumber\\
  &+& \sum_k^N \sum_{j>k}^N  \int_\infty d\bar{\mathbf{r}}_{j,k}  \int_\Omega d\mathbf{r}_k \int_\Omega d\mathbf{r}_j  \frac{\partial R^2}{\partial t} U_{j,k}.
\label{power1}
\end{eqnarray}
Equation \eref{power1} already contains the information we are looking for. 
Nonetheless, in order to understand the physical meaning of its main building-blocks, in the following we will manipulate it in order to get a more comprehensive result. 

By introducing the mentioned hydrodynamic form of the wavefunction $\Psi=Re^{iS}$ into Eq.~\eref{schrodinger} and then separating into imaginary and real parts, one gets respectively a continuity equation for the full probability density:
\begin{equation}
 \frac{\partial R^2}{\partial t} + \sum_j^N \nabla_j \cdot \mathbf{J}_j = 0,
\label{continuity} 
\end{equation}
and a quantum Hamilton-Jacobi-like equation for the full phase~ \cite{holland,bohm1,Wyatt,bohm2}:
\begin{equation}
 \frac{\partial S}{\partial t} + U + \sum_j^N Q_j + \sum_j^N \frac{1}{2}(\nabla_j S)^2 = 0. 
\label{QHJ} 
\end{equation}
Introducing Eqs.~\eref{continuity} and \eref{QHJ} into Eq.~\eref{power1} we find:
\begin{eqnarray}
 P_\Omega(t) = &-& \sum_k^N \int_\infty d\bar{\mathbf{r}}_k \int_\Omega d\mathbf{r}_k \nabla_k \left(R^2(\nabla_k S)Q_k) + R^2(\nabla_k S) \frac{(\nabla_k S)^2}{2} \right) \nonumber\\
               &+& \sum_k^N \int_\infty d\bar{\mathbf{r}}_k \int_\Omega d\mathbf{r}_k \sum_\xi^N R^2(\nabla_\xi S)(\nabla_\xi Q_k) \nonumber\\
               &+& \sum_k^N \int_\infty d\bar{\mathbf{r}}_k \int_\Omega d\mathbf{r}_k R^2 \frac{dQ}{dt} \nonumber\\
               &-& \sum_k^N \int_\infty d\bar{\mathbf{r}}_k \int_\Omega d\mathbf{r}_k R^2(\nabla_k S) (\nabla_k U_{ext}) \nonumber\\
               &-& \sum_k^N \int_\infty d\bar{\mathbf{r}}_k \int_\Omega d\mathbf{r}_k R^2(\nabla_k S) \left( \nabla_k \sum_j^N\sum_{\xi > j}^N U_{j,\xi} \right) \nonumber\\
               &-& \sum_k^N \int_\infty d\bar{\mathbf{r}}_k \int_\Omega d\mathbf{r}_k R^2(\nabla_k S) \left( \nabla_k \sum_\xi^N  Q_\xi \right) \nonumber\\
               &-& \sum_k^N \sum_{j>k}^N  \int_\infty d\bar{\mathbf{r}}_{j,k}  \int_\Omega d\mathbf{r}_k \int_\Omega d\mathbf{r}_j \sum_\xi^N (\nabla_\xi \mathbf{J}_\xi) U_{k,j}.
\label{power2}        
\end{eqnarray}
The above equation can be greatly simplified if one defines an external force,
\begin{eqnarray}
 \mathbf{F}^{ext}(\mathbf{r}_k,t) &=& -\nabla_k U_{ext}(\mathbf{r}_k,t), 
\end{eqnarray}
the $k$-th component of the Coulomb force,
\begin{eqnarray}
\mathbf{F}_k^{cou}(\mathbf{r}) &=& - \nabla_k \sum_j^N\sum_{\xi > j}^N U_{j,\xi}, 
\end{eqnarray}
and the $k$-th component of a quantum force,
\begin{eqnarray}
 \mathbf{F}_k^{qua}(\mathbf{r},t) &=& - \nabla_k  Q(\mathbf{r},t),
\end{eqnarray}
that arises in the context of a hydrodynamical representation of the wavefunction due to the presence of the (full) quantum potential $Q(\mathbf{r},t) = \sum_\xi^N  Q_\xi(\mathbf{r},t)$~\cite{bohm1,bohm2}.
Taking into account these force definitions and rewriting the first term on the r.h.s of Eq. (\ref{power2}) by using the Gauss theorem, we can write:
\begin{eqnarray}
 P_\Omega(t) = &-& \sum_k^N \int_\infty d\bar{\mathbf{r}}_k \left[ \mathbf{u}_x \cdot \mathbf{J}_k \left( Q_k + \frac{(\nabla_k S)^2}{2} \right) \right]_{S1}^{S4} \nonumber\\
            &+& \sum_k^N \int_\infty d\bar{\mathbf{r}}_k \int_\Omega d\mathbf{r}_k \sum_\xi^N \mathbf{J}_\xi \cdot (\nabla_\xi Q_k) \nonumber\\
            &+& \sum_k^N \int_\infty d\bar{\mathbf{r}}_k \int_\Omega d\mathbf{r}_k R^2 \frac{dQ}{dt} \nonumber\\
            &+& \sum_k^N \int_\infty d\bar{\mathbf{r}}_k \int_\Omega d\mathbf{r}_k \mathbf{J}_k \cdot (\mathbf{F}^{ext} + \mathbf{F}_k^{cou} + \mathbf{F}_k^{qua}) \nonumber\\
            &-& \sum_k^N \sum_{j>k}^N  \int_\infty d\bar{\mathbf{r}}_{j,k}  \int_\Omega d\mathbf{r}_k \int_\Omega d\mathbf{r}_j \sum_\xi^N \nabla_\xi (\mathbf{J}_\xi U_{k,j} ) \nonumber\\
            &+& \sum_k^N \sum_{j>k}^N  \int_\infty d\bar{\mathbf{r}}_{j,k}  \int_\Omega d\mathbf{r}_k \int_\Omega d\mathbf{r}_j \sum_\xi^N \mathbf{J}_\xi \cdot \nabla_\xi U_{k,j},
\end{eqnarray}
where we have used the following definition $[f(\mathbf{r})]_{S1}^{S4} = \int dy_k dz_k f(\mathbf{r})|_{x_k=x_{S4}} - \int dy_k dz_k f(\mathbf{r})|_{x_k=x_{S1}}$, and $\mathbf{u}_{x} \cdot \mathbf{J}_{k}$ is the longitudinal component 
of the $k$-th component of the current probability density.
Furthermore, we can use the following two identities. 
First,
\begin{eqnarray}
 \sum_k^N \int_\infty d\bar{\mathbf{r}}_k \int_\Omega d\mathbf{r}_k \mathbf{J}_k \cdot \mathbf{F}_k^{cou}   
       +  \sum_k^N \sum_{j>k}^N  \int_\infty d\bar{\mathbf{r}}_{j,k}  \int_\Omega d\mathbf{r}_k \int_\Omega d\mathbf{r}_j \sum_\xi^N \mathbf{J}_\xi \cdot \nabla_\xi U_{k,j}   \nonumber\\
     = \sum_k^N \int_\infty d\bar{\mathbf{r}}_k \int_\Omega d\mathbf{r}_k \mathbf{J}_k \cdot \mathbf{F}_k^{cou,ext},
\label{id1}
\end{eqnarray}
where we have defined the Coulombic force generated by all electrons outside the volume $\Omega$ on the $k$-th electron as $\mathbf{F}_k^{cou,ext}(\mathbf{r}) = - \nabla_k  \sum_{j\neq k} U_{k,j}^{ext}$, 
being $U_{k,j}^{ext} = U_{k,j}\theta(x_j - x_{S1})\theta(x_{S4} - x_j)$.
And second:
\begin{eqnarray}
\fl
 \sum_k^N \sum_{j>k}^N  \int_\infty d\bar{\mathbf{r}}_{j,k}  \int_\Omega d\mathbf{r}_k \int_\Omega d\mathbf{r}_j \sum_\xi^N \nabla_\xi (\mathbf{J}_\xi U_{k,j} ) 
    =  \sum_k^N  \int_\infty d\bar{\mathbf{r}}_{k}  \left[  \mathbf{u}_x \cdot \mathbf{J}_k \left(  \sum_{j\neq k} U_{k,j}^{int}  \right)  \right]_{S1}^{S4} 
\label{id2}
\end{eqnarray}
where we have defined the internal potential energy $U_{k,j}^{int} = U_{k,j}\theta(x_{S1} - x_j)\theta(x_j - x_{S4})$.
Using Eqs.~\eref{id1} and \eref{id2} we already attain the second main result of this work:
\begin{eqnarray}
  P_\Omega(t) &=& \sum_k^N \int_\infty d\bar{\mathbf{r}}_k \int_\Omega d\mathbf{r}_k    \mathbf{J}_k \cdot ( \mathbf{F}^{ext} + \mathbf{F}_k^{cou,ext} + \mathbf{F}_k^{qua}) \nonumber\\
         &-& \sum_k^N \int_\infty d\bar{\mathbf{r}}_k  \Big[ \mathbf{u}_{x} \cdot \mathbf{J}_{k} \Big( Q_k  + \frac{(\nabla_k S)^2}{2} + \sum_{j\neq k}^N U_{k,j}^{int} \Big) \Big]_{S1}^{S4} \nonumber\\  
         &+& \sum_k^N \int_\infty d\bar{\mathbf{r}}_k \int_\Omega d\mathbf{r}_k  R^2\frac{\partial Q_k}{\partial t}  \nonumber\\
         &+& \sum_k^N \int_\infty d\bar{\mathbf{r}}_k \int_\Omega d\mathbf{r}_k  \left( \sum_{j}^N \mathbf{J}_j \cdot \nabla_j   Q_k   \right).
\label{local_power}        
\end{eqnarray}
The power transferred by an external (effective) force to an ensemble of electrons can be locally resolved according to Eq.~\eref{local_power}. 
In the following subsection we discuss term by term the physical soundness of the above result.

\subsection{On the physical meaning of the expectation value of local power}
First of all, and as we did for the local energy in Eq.~\eref{local_energy}, we perform a first test of validity for Eq.~(\ref{local_power}) by taking the (previously introduced) classical limit \cite{Rosen}. 
By simply setting the quantum potential $Q_k$ to zero we get:
\begin{eqnarray}
  P_{\Omega,class}(t) &=& \sum_k^N \int_\infty d\bar{\mathbf{r}}_k \int_\Omega d\mathbf{r}_k    \mathbf{J}_{k,class} \cdot ( \mathbf{F}^{ext} + \mathbf{F}_k^{cou,ext}) \nonumber\\
         &-& \sum_k^N \int_\infty d\bar{\mathbf{r}}_k  \Big[ \mathbf{u}_{x} \cdot \mathbf{J}_{k,class} \Big( \frac{(\nabla_k S_{class})^2}{2} + \sum_{j\neq k}^N U_{k,j}^{int} \Big) \Big]_{S1}^{S4},
\label{local_power_classic}        
\end{eqnarray}
where $\mathbf{J}_{k,class}=R_{class}\nabla_k S_{class}$ according to the classical wavefunction $\Psi_{class} = R_{class}e^{iS_{class}}$~\cite{bohm1,Rosen}. The first term in Eq.~(\ref{local_power_classic}) 
corresponds to the standard definition of power as the product of the current by the external force, $\mathbf{J}_k \cdot \mathbf{F}^{ext}$~\cite{yang2}. 
Here, however, as already discussed in Sec.~\ref{local_pot}, particles outside $\Omega$ are effectively contributing to the external field in the form of $\mathbf{J}_k \cdot \mathbf{F}_k^{cou,ext}$. 
The second term in Eq.~\eref{local_power_classic} simply arises because the volume $\Omega$ is finite, i.e. energy can flow across the surfaces $S1$ and $S4$ along with the electrons.

In our discussion of Eq.~\eref{local_power}, we are now left with all pure quantum contributions to the local power. 
Let us first focus on the current-force correlation $\mathbf{J}_k \cdot \mathbf{F}_k^{qua}$ included in the first term of Eq.~(\ref{local_power}).
This term accounts for the fact that time variations of the kinetic energy in the volume $\Omega$ can be also due to the work done by a force with no classical counterpart. 
More specifically, according to our hydrodynamical language \cite{holland,bohm1,Wyatt,bohm2},  the quantum force $\mathbf{F}_k^{qua}$ is responsible for those changes in the current density that are not directly associated 
to the action of an electrical force~\cite{bohm5}. 
Consider, e.g., an electron impinging into a double barrier structure. 
It can happen that at later times the electron occupies a resonant state of the double barrier, and thus its associated current density decreases significantly. 
Such a damping effect is due (not because of the action of an external or Coulombic force, but) because of the quantum force $\mathbf{F}_k^{qua}$, which accounts for the confining effects within the double barrier~\cite{bohm3,bohm4,bohm6}.  
Notice that, due to the finite nature of the volume $\Omega$, in addition to this quantum current-force correlations, the quantum force is accompanied by the flux of quantum potential energy  
across the surfaces $x = S4$ and $x=S1$ (which is included in the second term of Eq.~\eref{local_power}). 
Finally, the third and fourth terms in Eq.~\eref{local_power} arise because of the explicit time variations of the quantum potential energy density $R^2Q_k$ in the volume $\Omega$. 
While the third term accounts for time variations of the quantum potential $Q_k$, the fourth term rates the change on the quantum potential energy due to the 
rearrangement of all electronic positions.

In the remaining part of the manuscript we consider a clarifying limit of our expression in Eq.~\eref{local_power}. 
We want to take the limit, $\Omega \rightarrow \infty$, where we implicitly assume that the extension of the support of the many-particle wavefunction is finite (for $\mathbf{r}^2 R^2 \to 0$ as $\mathbf{r} \to \infty$).  
In order our local energy definition in Eq.~\eref{local_energy} (and hence our finding in Eq.~\eref{local_power}) to be correct, in this limit we should recover Eq.~\eref{total_power}~\footnote{
Note that the closed system limit can be equivalently taken in two different ways: either we place the surfaces $S4$ and $S1$ far enough such that the full wavefunction vanishes there, 
or we make $x_{S4} = x_{S1}$ (for periodic systems)}. We will mathematically denote this close-system limit as $\Omega \to \infty$. 
Equation (\ref{local_power}) now reads:
\begin{eqnarray}
  P(t) &=& \sum_k^N \int_\infty d\mathbf{r}  \mathbf{J}_k \cdot ( \mathbf{F}^{ext} + \mathbf{F}_k^{cou,ext} + \mathbf{F}_k^{qua}) \nonumber\\
         &-& \sum_k^N \int_\infty d\bar{\mathbf{r}}_k  \Big[ \mathbf{u}_{x} \cdot \mathbf{J}_{k} \Big( Q_k  + \frac{(\nabla_k S)^2}{2} + \sum_{j\neq k}^N U_{k,j}^{int} \Big) \Big]_{S1}^{S4} \nonumber\\  
         &+& \sum_k^N \int_\infty d\mathbf{r}  R^2\frac{\partial Q_k}{\partial t}  
         + \sum_{k,j}^N \int_\infty d\mathbf{r} \left( \mathbf{J}_j \cdot \nabla_j   Q_k   \right).
\label{local_power_lim1}        
\end{eqnarray}
The second term in Eq. (\ref{local_power_lim1}) simply vanishes because $\mathbf{r}^2 R^2 \to 0$ as $\mathbf{r} \to \infty$. In addition, the following three equalities hold:
\begin{eqnarray}
 \sum_k^N \int_\infty d{\mathbf{r}}  \mathbf{J}_k \cdot \mathbf{F}_k^{qua} &=&  -\sum_{k,j}^N \int_\infty d{\mathbf{r}} \mathbf{J}_k \cdot \nabla_k Q_j \nonumber\\
     &=& -\sum_{k,j}^N \int_\infty d{\mathbf{r}} \mathbf{J}_j \cdot \nabla_j Q_k,
\label{equality_1}     
\end{eqnarray}
and
\begin{eqnarray}
 \sum_{k}^N \int_\infty d{\mathbf{r}}  \mathbf{J}_k \cdot \mathbf{F}_k^{cou,ext} &=&  0,
\label{equality_2}  
\end{eqnarray}
and
\begin{eqnarray}
  \int_\infty d{\mathbf{r}} R^2\frac{\partial Q_k}{\partial t} &=& 
          - \frac{1}{2} \int_\infty d{\mathbf{r}} R^2 \frac{\partial}{\partial t} \Big(\frac{\nabla_k^2 R}{R}\Big) \nonumber\\
      &=& - \frac{1}{2} \int_\infty d{\mathbf{r}} \nabla_k \Big(  R\nabla_k\frac{\partial R}{\partial t}   -  \frac{\partial R}{\partial t} \nabla_k R   \Big) \nonumber\\
      &=& 0.
\label{equality_3}       
\end{eqnarray}
Introducing the above three equalities into Eq.~\eref{local_power_lim1}, the expectation value of the total power finally reduces to:
\begin{eqnarray}
  P(t) = \sum_k^N \int_\infty d\mathbf{r}    \mathbf{J}_k \cdot \mathbf{F}^{ext}.
\label{local_power_closed2}        
\end{eqnarray}
Therefore, for the (full) closed quantum system we recover the classical-like expression already found in Eq.~\eref{total_power}. 
Equation (\ref{local_power_closed2}) does only include the single-particle classical force $\mathbf{F}^{ext}$. 
Many-body effects are now only implicit in the many-body current components $\mathbf{J}_k$, and non-local effects originating from the time derivative of the quantum potential $Q_k$ just canceled out.
Therefore, the above limit demonstrates that any explicit many-body or quantum signature found in Eq.~\eref{local_power} originates only because of the openness of the volume $\Omega$.

\section{Conclusions}
In this work we derived exact expressions for the spatial distribution of energy and power for quantum systems consisting of $N$ (interacting) spinless electrons in an effective (driving or dissipative) electric field.
Expectation values of the local energy and the local power, respectively in Eqs.~\eref{local_energy} and \eref{local_power}, are the main results of this work. 
Our final results were written in terms of an arbitrarily small volume $\Omega$, and interpreted from the quantum hydrodynamic point of view. 
Such a representation was proved to be helpful in the understanding of the main building-blocks of Eqs.~\eref{local_energy} and \eref{local_power}. 

The expectation value of the local (internal) energy in Eq.~\eref{local_energy} comprises two terms, viz., the kinetic and the potential energy contributions, which are respectively represented by the local operators derived
in Eqs.~\eref{kinetic_energy_open} and \eref{Coulombint}.
The expectation value of the local kinetic energy in Eq.~\eref{kineticintpo} can be further decomposed into a classical-like term, which is directly related to the current probability density, 
and a non-classical term, which is linked to the so-called quantum potential. 
The expectation value of the local potential energy in Eq.~\eref{Coulombint3} is the sum of contributions associated to pairs of electrons that (both) lay in the volume $\Omega$. 

The expectation value of local power in Eq.~\eref{local_power}, was found to consist of four main terms.
Classical-like contributions, explicitly written in Eq.~\eref{local_power_classic}, are represented by single-particle and many-body current-force correlations respectively of the form $\mathbf{J}_k \cdot \mathbf{F}^{ext}$ and $\mathbf{J}_k \cdot \mathbf{F}^{cou,ext}$, 
and accompanied by the flow of (classical) energy across the surfaces $S4$ and $S1$.
Explicit quantum contributions to the local power were proved to be directly linked to the quantum potential concept.
In this respect, the quantum potential contributes to the local power not only in the form of purely quantum current-force correlations $\mathbf{J}_k \cdot \mathbf{F}^{qua}$ and the inflow/outflow of quantum energy across the borders of 
the volume $\Omega$, but also, non-locally, through time variations of its associated energy density, which lead to the last two terms in Eq.~\eref{local_power}.

The soundness of our results in Eqs.~\eref{local_energy} and \eref{local_power} was first proven in relation to the classical limit $Q\to 0$, which lead to Eqs.~\eref{local_energy_classic} and \eref{local_power_classic} respectively.
Might be more intriguing was the closed system limit $\Omega\to\infty$. In particular, this limit greatly simplified the final expression for the total power by reducing it to the well-known expression in Eq.~\eref{total_power} (or equivalently Eq.~\eref{local_power_closed2}). 
In this respect, an interesting conclusion of our work is that both quantum and many-body features, explicitly manifested in the expression for the local power in Eq.~\eref{local_power}, cannot be distinguished anymore 
when power is integrated allover the physical space.
Non-locality and many-body correlations arising in Eq. \eref{local_power} will be numerically studied in future works for low-dimensional system of interest,
targeting low-power and energy harvesting nanoscale devices.

We want to finally stress that neither irreversible processes nor thermodynamical arguments were considered in this work.
Nonetheless, local power for a quantum system has shown to comprehend intriguing features that we expect to concur also in irreversible (dissipative) quantum processes.

\ack{
G. A acknowledges financial support from the Beatriu de Pin\'os program through the Project: 2014 BP-B 00244. 
F. L. T acknowledges support from the DOE under grant DE-FG02-05ER46204.
X. O acknowledges support from the \lq\lq{}Ministerio de Ciencia e Innovaci\'{o}n\rq\rq{} through the Spanish Project TEC2012-31330, Generalitat de Catalunya (2014 SGR-384) 
and the Grant agreement no: 604391 of the Flagship initiative  \lq\lq{}Graphene-Based Revolutions in ICT and Beyond\rq\rq{}}.


\section*{References}

\begin{thebibliography}{99}

%\bibitem{laser_cooling}
%Shuman E S, Barry J F, and DeMille D 2010 {\it Nature} {\bf 467} 820;

\bibitem{thermo1}
Solinas P, Averin D V, and Pekola J P 2013 {\it Physical Review B} {\bf 87} 060508;
\bibitem{thermo2}
Salmilehto J, Solinas P, and M\"ott\"onen M 2014 {\it Physical Review E} {\bf 89} 052128;
\bibitem{thermo3}
Venkatesh B P, Watanabe G, and Talkner P 2015 arXiv preprint arXiv:1503.03228;
\bibitem{thermo4}
Talkner P, and H\"anggi P 2015 arXiv preprint arXiv:1512.02516;
\bibitem{thermo5}
Campisi M, H\"anggi P, and Talkner P 2011 {\it Reviews of Modern Physics} {\bf 83} 771;



\bibitem{electronics1}
Tian B, Zheng X, Kempa T J, Fang Y, Yu N, Yu G, Huang J, Lieber C M 2007 {\it Nature} {\bf 449} 885.
\bibitem{electronics2}
Chang H-Y, et al. 2013 {\it Acs Nano} {\bf 7} 5446.
\bibitem{electronics3}
Wang H, et al. 2012 {\it Nano letters} {\bf 12} 4674.
\bibitem{electronics4}
Gu L, et al. 2012 {\it Nano letters} {\bf 13} 91.
\bibitem{electronics5}
Qin Y, Xudong W, and Zhong L W 2008 {\it Nature} {\bf 451} 809.



\bibitem{yang1}
Yang K-H 1983 {\it Journal of Physics A: Mathematical and General} {\bf 16} 935.


\bibitem{cooling}
Phillips W D 1998 {\it Reviews of Modern Physics} {\bf 70} 721.


\bibitem{local_electrons1}
D'Agosta R, and Di Ventra M 2008 {\it Journal of Physics: Condensed Matter} {\bf 20} 374102.
\bibitem{local_electrons2}
D'Agosta R, Sai N, and Di Ventra M 2006 {\it Nano letters} {\bf 6} 2935.
\bibitem{local_electrons3}
Tsutsui M, Kawai T, and Taniguchi M 2012 {\it Scientific reports} {\bf 2}.


\bibitem{yang_classic}
Kobe D H, and Yang K-H 1987 {\it European Journal of Physics} {\bf 8} 236.

\bibitem{measuring}
Ioffe Zvi, et al. 2008 {\it Nature nanotechnology} {\bf 3} 727.

\bibitem{yang2}
Kobe D H, Wen E C-T, and Yang K-H  1982 {\it Physical Review D} {\bf 26} 1927.

\bibitem{cohen}
Cohen-Tannoudji C, Diu B, and Laloe F 2009 {\it Quantum mechanics. Vol. 1} Willey.

\bibitem{holland}
Holland P R 1995 {\it The quantum theory of motion: an account of the de Broglie-Bohm causal interpretation of quantum mechanics}. Cambridge university press.

\bibitem{bohm1}
Oriols X, and Mompart J, eds. 2012 {\it Applied Bohmian mechanics: From nanoscale systems to cosmology} (Singapore: Pan Stanford Publishing).

\bibitem{Wyatt}
Wyatt R 2005 2012 {\it Quantum Dynamics with trajectories: Introduction to quantum hydrodynamics} (USA: Spinrger).

\bibitem{bohm2}
Benseny A, Albareda G, Sanz A S, Mompart J, and Oriols X 2014 {\it The European Physical Journal D} {\bf 68} 1.

\bibitem{extra}
The definition in Eq.~\eref{Coulombint} can be also motivated in a very different way.  Remind that it is the aim of our local energy operator to rigorously define the concept of internal energy associated to the volume $\Omega$. Consider the limit of $\Omega \to 0$. Since $\Omega$ is now infinitesimal, at most one electron can sit in it. 
Then, if the full probability density at $\Omega$ is zero, then the associated potential energy is obviously zero. However, if one electron falls in $\Omega$, then the question of what is the internal energy associated to the volume $\Omega$ now turns into the question what is the internal energy of such an electron.  As an indivisible particle, the internal energy of an electron can only be its kinetic energy, and hence the potential energy contribution must be zero. This example highlights the conceptual validity of the operator in Eq.~\eref{Coulombint}, which in the limit $\Omega\to 0$ reduces to zero, i.e. 
the internal energy is only composed of kinetic energy.

\bibitem{Rosen}
Rosen N 1964 {\it  American Journal of Physics}  {\bf 32}  377.


\bibitem{bohm3}
Albareda G, Marian D, Benali A, Yaro S, Zangh\`i N, and Oriols X 2013 {\it Journal of Computational Electronics} {\bf 12} 405.

\bibitem{bohm4}
Albareda G, Su\~n\'e J, and Oriols X 2009 {\it Physical Review B} {\bf 79} 1.

\bibitem{bohm5}
Albareda G, Traversa F L, Benali A, and Oriols X 2012 {\it Fluctuation and Noise Letters} {\bf 11} 1242008.

\bibitem{bohm6}
Traversa F L, et al. 2011 {\it IEEE Transactions on Electron Devices} {\bf 58} 2104.

% \bibitem{bohm6}
% Benali A, Traversa F L, Albareda G, Alarcon A, Aghoutane M, and Oriols X 2012 {\it Fluctuation and Noise Letters} {\bf 11} 1241002.
% 
% \bibitem{bohm7}
% Benali A, Traversa F L, Albareda G, Aghoutane M, and Oriols X 2013 {\it Applied Physics Letters} {\bf 102} 173506.




% 
% A. C. Smith, J. F. Janak, and R. B. Adler, Electronic Conduction in Solids. New York, NY, USA: McGraw-Hill, 1967
% 
% IEEE ELECTRON DEVICE LETTERS, VOL. 36, NO. 1, JANUARY 2015
% Study of Local Power Dissipation in Ultrascaled
% Silicon Nanowire FETs
% Antonio Martinez, John R. Barker, Manuel Aldegunde, and Raul Valin
% 
% Lake, Roger, and Supriyo Datta. "Energy balance and heat exchange in mesoscopic systems." Physical Review B 46.8 (1992): 4757.
% 
% Energy Dissipation and Transport in Nanoscale Devices Pop

% 
% 
% 
% \bibitem{ITRS}
% International Technology Roadmap for Semiconductors (2010 Update) http://www.itrs.net 










\end{thebibliography}
\end{document}